\documentclass{article}
\usepackage[preprint]{spconfa4}
\usepackage{amsmath,graphicx}
\usepackage[style=ieee,giveninits=true,url=false,doi=false,isbn=false]{biblatex}
\AtEveryBibitem{\clearlist{language}}
\AtEveryBibitem{\clearfield{eprint}}
\AtEveryBibitem{\clearfield{note}}
\usepackage[nolist]{acronym}
\usepackage{siunitx}
\usepackage{bbold}
\usepackage{empheq}
\usepackage[extreme,bibliography=normal,title=normal,sections=normal,margins=normal]{savetrees}
\usepackage{booktabs}
\usepackage{tabularx}
\usepackage[hidelinks]{hyperref}
\usepackage[ruled, lined, linesnumbered, commentsnumbered, noend]{algorithm2e}
\usepackage{subcaption}
\usepackage{svg}

\begin{acronym}
	\acro{STFT}{short-time Fourier transform}
	\acro{SNR}{signal-to-noise ratio}
	\acro{PESQ}{perceptual evaluation of speech quality}
	\acro{STOI}{short-time objective intelligibility}
	\acro{MVDR}{minimum variance distortionless response}
	\acro{RTF}{relative transfer function}
	\acro{DNN}{deep neural network}
	\acro{CW}{covariance whitening}
\end{acronym}

\title{Dictionary-Based Fusion of Contact and Acoustic Microphones for Wind Noise Reduction}
\name{Marvin Tammen$^{1,2,*}$, Xilin Li$^1$, Simon Doclo$^2$, Lalin Theverapperuma$^1$\thanks{$^*$This work was performed while Marvin Tammen was an intern with Meta Reality Labs Research.}}
\address{$^1$Meta Reality Labs Research, California, USA\\
	$^2$University of Oldenburg and Cluster of Excellence Hearing4all, Oldenburg, Germany}
\DeclareMathOperator{\argmin}{argmin}

\DeclareMathOperator{\prox}{prox}
\newcommand{\norm}[1]{\left\lVert#1\right\rVert}

\begin{document}
\ninept
\maketitle
\begin{abstract}
In mobile speech communication applications, wind noise can lead to a severe reduction of speech quality and intelligibility.
Since the performance of speech enhancement algorithms using acoustic microphones tends to substantially degrade in extremely challenging scenarios, auxiliary sensors such as contact microphones can be used.
Although contact microphones offer a much lower recorded wind noise level, they come at the cost of speech distortion and additional noise components.
Aiming at exploiting the advantages of acoustic and contact microphones for wind noise reduction, in this paper we propose to extend conventional single-microphone dictionary-based speech enhancement approaches by simultaneously modeling the acoustic and contact microphone signals.
We propose to train a single speech dictionary and two noise dictionaries and use a relative transfer function to model the relationship between the speech components at the microphones.
Simulation results show that the proposed approach yields improvements in both speech quality and intelligibility compared to several baseline approaches, most notably approaches using only the contact microphones or only the acoustic microphone.
\end{abstract}
\section{Introduction}
In many mobile speech communication scenarios, ambient noise such as wind noise can lead to a severe reduction of speech intelligibility and quality in the recorded microphone signals.
To increase speech intelligibility in these scenarios, speech enhancement algorithms using one or more acoustic microphones can be applied.
The performance of such algorithms tends to decrease in extremely challenging scenarios, as model assumptions become less valid in case of model-based algorithms or the mismatch between training and test data becomes too large in case of supervised learning-based algorithms.

To increase speech enhancement performance in such challenging scenarios, it has been proposed to include auxiliary information to help separate the target speaker from ambient noise, e.g., video signals~\cite{michelsanti_overview_2021} or signals from more noise-resistant microphones such as bone conduction microphones~\cite{shin_survey_2012,zhou_robust_2021}.
In particular, contact microphones can be exploited for wind noise reduction, since they offer the advantage of much lower wind noise levels at the disadvantage of speech distortion and additional noise components.
To alleviate these disadvantages, algorithms performing color correction or bandwidth extension~\cite{bouserhal_-ear_2017} can be utilized.
These algorithms exploit the fact that the \ac{SNR} at the contact microphone is often higher than at the acoustic microphone and try to compensate for the speech distortion by equalizing the frequency response of the contact microphone or restoring missing harmonic components.
However, these algorithms typically do not make use of the complementary information contained in the acoustic microphone.
Thus, an important question is how to combine conventional acoustic microphones with contact microphones, aiming at combining the advantages of both while overcoming their disadvantages.
A recent popular approach to fuse information from different sensors or modalities is to leverage big data, e.g., by training \acp{DNN}~\cite{michelsanti_overview_2021,gao_survey_2020,yu_time-domain_2020}.
\acp{DNN} perform a non-linear mapping from the input space to the output space and can perform quite well, provided that a sufficient amount of representative training data is available.
However, due to their highly complex nature they typically act as a black box, rendering interpretations of the learned models rather difficult.

Alternatively, prior knowledge about the relationship between the sensors as well as the typically encountered signals can be exploited, e.g., in the form of statistical models~\cite{gustafsson_statistical_2010,shin_priori_2015} or dictionary-based sparse signal representation~\cite{singh_multimodal_2020}.
While dictionary-based speech enhancement approaches~\cite{jafari_fast_2011,sigg_learning_2012,sigg_speech_2012,ji_speech_2019} also rely on the availability of suitable training data, the required dataset size is typically in the range of minutes and thus much smaller than for \acp{DNN}.
In addition, the interpretability of trained dictionaries is much higher.

Aiming at exploiting the advantages of acoustic and contact microphones for wind noise reduction, in this paper we propose to extend conventional single-microphone dictionary-based speech enhancement approaches such as~\cite{sigg_speech_2012,ji_speech_2019} by simultaneously modeling the acoustic and contact microphone signals. 
More specifically, we propose to train a single speech dictionary and use a \acl{RTF} to model the relationship between the speech components at the acoustic and contact microphones.
Furthermore, to increase the speech and noise separation capability, we propose to use two noise dictionaries: one dictionary for representing typical noise at the acoustic microphone, and another dictionary for the contact microphone. 
Simulation results comprising recorded wind noise at different wind speeds and \acp{SNR} show improvements in both speech quality and intelligibility w.r.t. a number of baseline approaches, most notably approaches using only the contact microphone or only the acoustic microphone. %, as well as the color correction algorithm and a traditional \ac{MVDR} beamformer.

\section{Signal Model}
We consider an acoustic scenario containing two microphones with different characteristics, i.e., an acoustic and a contact microphone, recording a single speech source and wind noise, which is assumed to be additive.
The acoustic microphone, denoted by superscript $\circ^A$, exhibits a full-band frequency response and high susceptibility to wind noise, whereas the contact microphone, denoted by superscript $\circ^B$, captures band-limited speech while being less susceptible to wind noise and potentially recording additional independent noise (e.g., clicking noise caused by moving the mounting device).
In the \ac{STFT}-domain with frequency index $f$ and frame index $t$, the noisy microphone signals at both microphones can be written as
\begin{align}
	\label{eq: noisy}
	\begin{bmatrix}
		Y^A(f,t)\\Y^B(f,t)
	\end{bmatrix}
	=
	\begin{bmatrix}
		X^A(f,t) &+& N^A(f,t)\\X^B(f,t) &+& N^B(f,t)
	\end{bmatrix},
\end{align}
where $X^A(f,t)$, $X^B(f,t)$, $N^A(f,t)$, and $N^B(f,t)$ denote the target speech and noise components captured by the acoustic microphone $\circ^A$ and the contact microphone $\circ^B$, respectively.
Moreover, we assume that the speech components at both microphones are related using a \ac{RTF} $H(f)$, i.e., $X^B(f,t) = H(f) X^A(f,t)$.
%\begin{equation}
%	\label{eq:rtf}
%	X^B(f,t) = H(f) X^A(f,t).
%\end{equation}
It should be noted that the unknown \ac{RTF} is assumed to be time-invariant over $T$ time frames.
Aggregating all $F$ frequency bins and $T$ time frames, we denote the \ac{STFT} matrices of the noisy, target speech, and noise components as $\mathbf{Y}^A$, $\mathbf{Y}^B$, $\mathbf{X}^A$, $\mathbf{N}^A$, and $\mathbf{N}^B \in \mathbb{C}^{F \times T}$, respectively, such that \eqref{eq: noisy} can be written as
\begin{align}
	\label{eq: noisy and rtf, matrix}
	\begin{bmatrix}
		\mathbf{Y}^A\\\mathbf{Y}^B
	\end{bmatrix}
	=
	\begin{bmatrix}
		\mathbf{X}^A &+& \mathbf{N}^A\\\mathbf{H} \mathbf{X}^A &+& \mathbf{N}^B
	\end{bmatrix},
\end{align}
with $\mathbf{H} = \mathrm{diag}\left( \left\{ H(f) \right\} \right) \in \mathbb{C}^{F \times F}$ a diagonal matrix.
The goal of the proposed microphone fusion algorithm is to estimate the target speech component at the acoustic microphone $\mathbf{X}^A$ using the noisy acoustic and contact microphone signals $\mathbf{Y}^A$ and $\mathbf{Y}^B$.

\section{Dictionary-Based Microphone Fusion}
\subsection{Dictionary-Based Noise Reduction Using Single Microphone}
Previous studies~\cite{jafari_fast_2011,sun_single-channel_2021,sigg_speech_2012,kowalski_social_2013,mlynarski_sparse_2014} have shown that speech signals $\mathbf{X}^A$ can be compactly represented using an over-complete pre-trained speech dictionary $\mathbf{D}^A_X$ and a sparse code $\mathbf{C}^A_X$.
While most of these studies considered a real-valued dictionary and sparse code~\cite{jafari_fast_2011,sun_single-channel_2021,sigg_speech_2012,kowalski_social_2013}, similarly to \cite{mlynarski_sparse_2014} we consider a complex-valued dictionary $\mathbf{D}^A_X \in \mathbb{C}^{F \times N}$ and sparse code $\mathbf{C}^A_X \in \mathbb{C}^{N \times T}$, where $N > F$ denotes the number of dictionary atoms, i.e.,
\begin{align}
	\label{eq: dict representation speech}
	\mathbf{X}^A &\approx \widehat{\mathbf{X}}^A = \mathbf{D}^A_X \mathbf{C}^A_X,
\end{align}
with $\widehat{\circ}$ denoting the (dictionary-based) estimate of $\circ$.
The speech dictionary $\mathbf{D}^A_X$ is typically obtained in a training stage as the solution of the following non-convex optimization problem:
\begin{align}
	\label{eq: dict opt}
	\mathbf{D}^A_X, \ \mathbf{C}^A_X = \argmin_{\mathbf{D}_X, \mathbf{C}_X} \underbrace{\norm{\mathbf{X}^A - \mathbf{D}_X \mathbf{C}_X}^2_2}_{J_{\mathrm{rec}}} + \lambda \underbrace{\norm{\mathbf{C}_X}_1}_{J_{\mathrm{spa}}},
\end{align}
where $J_{\mathrm{rec}}$ represents the reconstruction cost incurred by the dictionary representation, and $J_{\mathrm{spa}}$ promotes sparsity of the code $\mathbf{C}^A_X$ or, in other words, encourages the usage of only a small number of dictionary atoms per time frame.
The parameter $\lambda$ enables to trade off between a more representative vs. a sparser representation.
To solve this non-convex optimization problem, typically methods which alternate between a dictionary update step and a sparse coding step are used~\cite{engan_multi-frame_2000,aharon_k-svd_2006,sigg_learning_2012}.

The idea of representing speech signals using a dictionary and a corresponding sparse code has been extended~\cite{sigg_speech_2012,sun_single-channel_2021,ji_speech_2019} to represent \emph{noisy} speech signals with known noise characteristics by considering both speech and noise dictionaries and sparse codes.
Similarly as for the speech component in \eqref{eq: dict representation speech}, the noise component can be represented as $\mathbf{N}^A \approx \widehat{\mathbf{N}}^A = \mathbf{D}^A_N \mathbf{C}^A_N$.
Assuming available speech and noise dictionaries $\mathbf{D}^A_X$ and $\mathbf{D}^A_N$, the speech and noise sparse codes are estimated from the noisy microphone signals as the solution of the following non-convex optimization problem:
\begin{align}
	&\argmin_{\mathbf{C}^A_X, \mathbf{C}^A_N} \norm{\mathbf{Y}^A - \widehat{\mathbf{Y}}^A}^2_2 + \lambda\norm{\begin{bmatrix} \mathbf{C}^A_X\\\mathbf{C}^A_N \end{bmatrix}}_1 \\
%	=&\argmin_{\mathbf{C}^A_X, \mathbf{C}^A_N} \norm{\mathbf{Y}^A - (\widehat{\mathbf{X}}^A + \widehat{\mathbf{N}}^A)}^2_2 + \lambda\norm{\begin{bmatrix} \mathbf{C}^A_X\\\mathbf{C}^A_N \end{bmatrix}}_1 \nonumber\\
	=&\argmin_{\mathbf{C}^A_X, \mathbf{C}^A_N} \norm{\mathbf{Y}^A - (\mathbf{D}^A_X \mathbf{C}^A_X + \mathbf{D}^A_N \mathbf{C}^A_N)}^2_2 + \lambda\norm{\begin{bmatrix} \mathbf{C}^A_X\\\mathbf{C}^A_N \end{bmatrix}}_1.\nonumber
\end{align}
Intuitively, components of the noisy microphone signal will be captured either by the speech dictionary, the noise dictionary, or not at all (e.g., Gaussian noise), depending on their coherence w.r.t. the respective dictionary.

\subsection{Dictionary-Based Noise Reduction Using Microphone Fusion}
\begin{figure}[t]
	\centering
	\includegraphics[width=\linewidth]{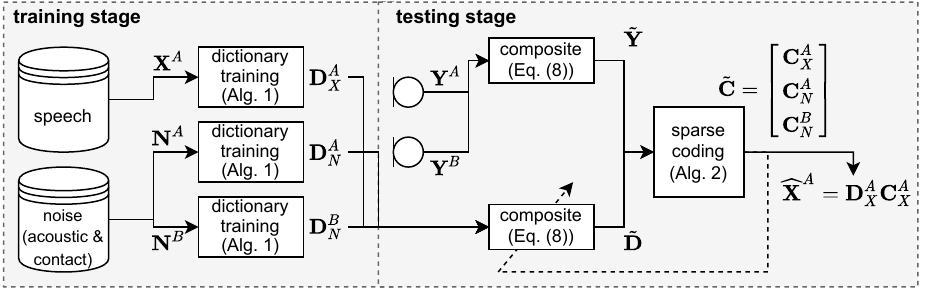}
	\caption{Block diagram of the proposed dictionary-based microphone fusion approach for wind noise reduction.
		In the training stage, an arbitrary clean speech dataset and a wind noise dataset are used to construct an acoustic speech and dictionary and acoustic and contact noise dictionaries.
		In the testing stage, using the composite dictionary and microphone signals, the speech and noise sparse codes are computed, such that the speech component can be estimated using the speech dictionary and its sparse code.}
	\label{fig:blockdiagram}
\end{figure}

Building upon the above single-microphone approach, we propose to simultaneously model both the acoustic and contact microphone signals by integrating the signal model \eqref{eq: noisy and rtf, matrix} into the dictionary training and sparse coding steps.
More specifically, we model the different noise components at the acoustic and contact microphone using two separate noise dictionaries $\mathbf{D}^A_N$ and $\mathbf{D}^B_N$, and we model the relationship of the speech components with an (unknown) \ac{RTF} $\mathbf{H}(f)$.
An overview of the proposed approach is depicted in Fig. \ref{fig:blockdiagram}.

Inserting the dictionary-based representation of the target speech and noise components in the signal model \eqref{eq: noisy and rtf, matrix} yields
\begin{empheq}{align}
	\label{eq: noisy and rtf, matrix, dict}
	\begin{bmatrix}
		\mathbf{Y}^A\\\mathbf{Y}^B
	\end{bmatrix}
	\approx \begin{bmatrix}
		\widehat{\mathbf{Y}}^A\\\widehat{\mathbf{Y}}^B
	\end{bmatrix}
	= \begin{bmatrix}
		\mathbf{D}^A_X \mathbf{C}^A_X &+& \mathbf{D}^A_N \mathbf{C}^A_N\\\mathbf{H} \mathbf{D}^A_X \mathbf{C}^A_X &+& \mathbf{D}^B_N \mathbf{C}^B_N
	\end{bmatrix}.
\end{empheq}

Based on \eqref{eq: noisy and rtf, matrix, dict}, the reconstruction cost in \eqref{eq: dict opt} can be extended to include both the acoustic and the contact microphone signals, i.e.,
\begin{align}
	\label{eq: cost rec}
	J_{\mathrm{rec}} &= \eta \norm{\mathbf{Y}^A - (\widehat{\mathbf{X}}^A + \widehat{\mathbf{N}}^A)}^2_2 + (1 - \eta) \norm{\mathbf{Y}^B - (\mathbf{H} \widehat{\mathbf{X}}^A + \widehat{\mathbf{N}}^B)}^2_2\nonumber\\
	&= \norm{\tilde{\mathbf{Y}} - \tilde{\mathbf{D}} \tilde{\mathbf{C}}}^2_2,
\end{align}
where the parameter $\eta$ allows to control the influence of each microphone, and
where the following composite quantities have been defined:
\begin{empheq}[left=\empheqlbrace]{align}
	\tilde{\mathbf{Y}} &= \begin{bmatrix}
		\sqrt{\eta} \mathbf{Y}^A\\\sqrt{1 - \eta} \mathbf{Y}^B
	\end{bmatrix}, \quad 
	\tilde{\mathbf{C}} = \begin{bmatrix}
		\mathbf{C}^A_X\\
		\mathbf{C}^A_N\\
		\mathbf{C}^B_N
	\end{bmatrix}\nonumber\\
	\tilde{\mathbf{D}} &= \begin{bmatrix}
		\sqrt{\eta} \mathbf{D}^A_X &\sqrt{\eta} \mathbf{D}^A_N &\mathbf{0}\\
		\sqrt{1 - \eta} \mathbf{H} \mathbf{D}^A_X &\mathbf{0} &\sqrt{1 - \eta} \mathbf{D}^B_N
	\end{bmatrix}.
	\label{eq: composites definition}
\end{empheq}

Similarly, the sparsity cost $J_{\mathrm{spa}}$ is extended to include the sparse codes of the speech component as well as the acoustic and contact noise components.
Thus, similarly to \eqref{eq: dict opt}, the total cost to be minimized can be written as
\begin{empheq}[box=\fbox]{align}
	\label{eq: composite cost}
	c = \norm{\tilde{\mathbf{Y}} - \tilde{\mathbf{D}} \tilde{\mathbf{C}}}^2_2 + \lambda \norm{\tilde{\mathbf{C}}}_1
\end{empheq}

\subsection{Dictionary Training}
The required dictionaries $\mathbf{D}^A_X$, $\mathbf{D}^A_N$ and $\mathbf{D}^B_N$ for the target speech and noise components are trained independently for each component by solving the optimization problem in \eqref{eq: dict opt}.
To this end, we use the alternating optimization procedure presented in Alg.~\ref{alg: dl}~\cite{parikh_proximal_2013,engan_multi-frame_2000}, where $\mathbf{S} \in \{ \mathbf{X}^A, \ \mathbf{N}^A, \ \mathbf{N}^B \}$.
After initializing the dictionary with random unit-normalized \ac{STFT} frames from the training data, the procedure alternates between optimizing \eqref{eq: dict opt} w.r.t. the sparse code and the dictionary.
\begin{algorithm}%TODO: mention alternative dictionary training algorithms and why we have chosen this one
	\DontPrintSemicolon
	\KwIn{\ac{STFT} matrix $\mathbf{S}$, sparsity weight $\lambda$, number of outer iterations $I_{\mathrm{dict}}$, inner iterations $I_{\mathrm{sc}}$, and atoms $N$}
	\KwOut{dictionary $\mathbf{D}$}
	init. $\mathbf{D}_0$ using randomly chosen \ac{STFT} frames, init. $\mathbf{C}_0 = \mathbf{0}$\;
%	init. $\mathbf{C}_0 = \mathbf{0}$\;
	\For{$i \in \{1, \ \dots, \ I_{\mathrm{dict}}\}$}{
		%		\tcp{compute sparse codes (Alg.~\ref{alg: sc})}
		$\mathbf{C}_i = \text{Alg. 2} \left( \mathbf{S}, \ \mathbf{D}_{i-1}, \ \mathbf{C}_{i-1}, \ \lambda, \ I_{\mathrm{sc}} \right)$\;
		%		\tcp{update dictionary~\cite{engan_multi-frame_2000}}
		$\mathbf{D}_i = \argmin_{\mathbf{D}} \norm{\mathbf{S} - \mathbf{D} \mathbf{C}_i}^2_2 + \lambda \norm{\mathbf{C}_i}_1 	= \mathbf{S} \mathbf{C}^{H}_i \left( \mathbf{C}_i \mathbf{C}^{H}_i \right)^{-1}$\;
		$\mathbf{D}_i \leftarrow$ normalize columns\;
	}
	return $\mathbf{D}_{I_{\mathrm{dict}}}$
	\caption{Dictionary Training}
	\label{alg: dl}
\end{algorithm}

For the sparse coding step at the $i$-th iteration, the optimization problem is given by
%\begin{equation}
%	\label{eq: sc}
%	\mathbf{C}_i = \argmin_{\mathbf{C}} \underbrace{\norm{\mathbf{S} - \mathbf{D}_{i-1} \mathbf{C}}^2_2}_{f\left( \mathbf{C} \right)} + \underbrace{\lambda \norm{\mathbf{C}}_1}_{g\left( \mathbf{C} \right)},
%\end{equation}
\begin{equation}
	\label{eq: sc}
	\mathbf{C}_i = \argmin_{\mathbf{C}} \norm{\mathbf{S} - \mathbf{D}_{i-1} \mathbf{C}}^2_2 + \lambda \norm{\mathbf{C}}_1.
\end{equation}
To solve this optimization problem, we use the accelerated proximal gradient method presented in Alg.~\ref{alg: sc}~\cite{parikh_proximal_2013}. %, which is described in detail in .

\begin{algorithm}
	\DontPrintSemicolon
	\KwIn{\ac{STFT} matrix $\mathbf{S}$, sparsity weight $\lambda$, number of inner iterations $I_{\mathrm{sc}}$, initial sparse code $\mathbf{C}_0$, dictionary $\mathbf{D}$}
	\KwOut{updated sparse code $\mathbf{C}$}
	$L =$ max eigenvalue $\left( \mathbf{D}^H \mathbf{D} \right)$\;
	$\mu = \frac{1}{L}$
	\tcp{Lipschitz constant-based step size}
	\For{$i \in \{1, \ \dots, \ I_{\mathrm{sc}}\}$}{
		$w_i = \frac{i}{i+1}$\tcp*{extrapolation step size}
%		\tcp{auxiliary variables:}
		$\boldsymbol{\Gamma}_i = \mathbf{C}_{i-1} + w_i \left( \mathbf{C}_{i-1} - \mathbf{C}_{i-2} \right)$\;
		$\mathbf{Z}_i = \boldsymbol{\Gamma}_i - \mu \mathbf{D}^H \left( \mathbf{D} \boldsymbol{\Gamma}_i - \mathbf{S} \right)$\;
		$\mathbf{C}_{i} = \prox_{\mu g} \left( \boldsymbol{\Gamma}_i - \mu \nabla f\left( \boldsymbol{\Gamma}_i \right) \right) = \left(1 - \frac{\mu \lambda}{\texttt{max} \left( |\mathbf{Z}_i|, \mu \lambda \right)}\right) \mathbf{Z}_i$\;
	}
	return $\mathbf{C}_{I_{\mathrm{sc}}}$
	\caption{Sparse Coding}
	\label{alg: sc}
\end{algorithm}

\subsection{Proposed Speech Enhancement Algorithm}
Assuming that the speech and noise dictionaries $\mathbf{D}^A_X$, $\mathbf{D}^A_N$, and $\mathbf{D}^B_N$ are available, the cost in \eqref{eq: composite cost} is a function of the composite sparse code $\tilde{\mathbf{C}}$ defined in \eqref{eq: composites definition} as well as the \ac{RTF} $\mathbf{H}$, which both need to be estimated. 
%While it may be viable in certain scenarios to assume that the \ac{RTF} between speech components at the acoustic and contact microphones is known, in this work we consider the \ac{RTF} as a quantity that needs to be estimated.
Thus, the proposed speech enhancement algorithm consists of solving the following optimization problem:
\begin{empheq}[box=\fbox]{align}
	\label{eq: problem se}
	\widehat{\tilde{\mathbf{C}}}, \ \widehat{\mathbf{H}} = \argmin_{\tilde{\mathbf{C}}, \ \mathbf{H}} \norm{\tilde{\mathbf{Y}} - \tilde{\mathbf{D}} \tilde{\mathbf{C}}}^2_2 + \lambda \norm{\tilde{\mathbf{C}}}_1
\end{empheq}
As no closed-form solution exists, the proposed speech enhancement algorithm presented in Alg.~\ref{alg: se} alternates between minimizing \eqref{eq: composite cost} w.r.t. the composite sparse code $\tilde{\mathbf{C}}$ and the \ac{RTF} $\mathbf{H}$.
%We initialize the \ac{RTF} estimate using the \ac{CW} method~\cite{markovich-golan_multichannel_2009}.
For sparse coding, we consider the same accelerated proximal gradient method as during dictionary training (cf. Alg.~\ref{alg: sc}).
For estimating the \ac{RTF} at the $i$-th iteration while fixing the sparse code to its current estimate, a closed-form solution can be obtained as
\begin{empheq}{align}
	\label{eq: problem rtf}
	\widehat{\mathbf{H}} &= \argmin_{\mathbf{H}} \norm{\tilde{\mathbf{Y}} - \tilde{\mathbf{D}}_i \widehat{\tilde{\mathbf{C}}}_i}^2_2 + \lambda \norm{\widehat{\tilde{\mathbf{C}}}_i}_1\\
	&= \left( \mathbf{Y}^B - \widehat{\mathbf{N}}^B_i\right) \widehat{\mathbf{X}}^{{A,H}}_i \left( \widehat{\mathbf{X}}^{{A}}_i \widehat{\mathbf{X}}^{{A,H}}_i \right)^{-1}.
\end{empheq}
Note that we only consider diagonal \ac{RTF} estimates $\widehat{\mathbf{H}}$ to avoid interactions between different frequency bins.
\begin{figure*}
	\centering
	\begin{subfigure}[t]{0.5\textwidth}
		\centering
		\includesvg[width=\linewidth]{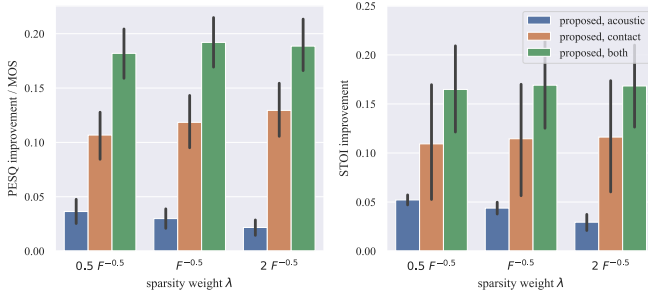}
		\caption{Proposed algorithms for different sparsity weights $\lambda$.}
		\label{fig:results: lambdas}
	\end{subfigure}%
	\begin{subfigure}[t]{0.5\textwidth}
		\centering
		\includesvg[width=\linewidth]{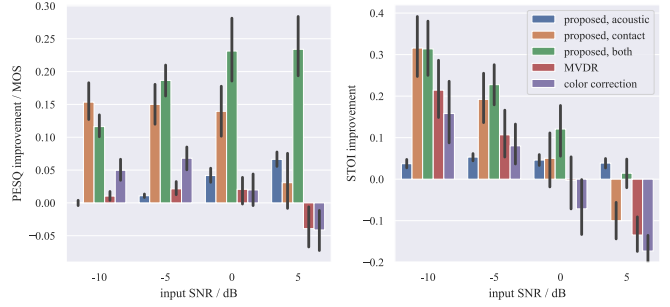}
		\caption{Proposed and baseline algorithms for different input \acp{SNR}.}
		\label{fig:results: baselines}
	\end{subfigure}
	\caption{Speech enhancement performance in terms of \acs{PESQ} and \acs{STOI} improvements w.r.t. the noisy acoustic microphone signal.}
	\label{fig:results}
\end{figure*}

Using the estimated composite sparse code $\widehat{\tilde{\mathbf{C}}}$, the speech component can be estimated by multiplying the estimated speech sparse code with the speech dictionaries, i.e., $\widehat{\mathbf{X}}^A = \mathbf{D}^A_X \widehat{\mathbf{C}}^A_X$.
%These estimates can be converted to the time-domain using an inverse \ac{STFT}.
\begin{algorithm}
	\DontPrintSemicolon
	\KwIn{noisy \ac{STFT} matrices $\mathbf{Y}^A, \mathbf{Y}^B$, composite dictionary $\tilde{\mathbf{D}}$, microphone weight $\eta$, sparsity weight~$\lambda$, number of outer iterations $I_{\mathrm{se}}$, inner iterations $I_{\mathrm{sc}}$}
	\KwOut{estimated speech \ac{STFT} matrix $\widehat{\mathbf{X}}^A$ and \ac{RTF} $\widehat{\mathbf{H}}$}
	construct $\tilde{\mathbf{Y}}$ using \eqref{eq: composites definition}\;
	init. $\widehat{\mathbf{H}}_0$ using \ac{CW}, init. $\widehat{\tilde{\mathbf{C}}}_0 = \mathbf{0}$\;
%	init. $\widehat{\tilde{\mathbf{C}}}_0 = \mathbf{0}$\;
	\For{$i \in \{1, \ \dots, \ I_{\mathrm{se}}\}$}{
		construct $\tilde{\mathbf{D}}_i$ using \eqref{eq: composites definition}\;
%		\tcp{compute sparse codes (Alg.~\ref{alg: sc})}
		$\widehat{\tilde{\mathbf{C}}}_{i} = \text{Alg. 2} \left( \tilde{\mathbf{Y}}, \ \tilde{\mathbf{D}}, \ \widehat{\tilde{\mathbf{C}}}_{i-1}, \ \lambda, \ I_{\mathrm{sc}} \right)$\;
		$\widehat{\mathbf{X}}^{A}_i = \begin{bmatrix} \mathbf{D}^A_X & \mathbf{0} & \mathbf{0} \end{bmatrix} \widehat{\tilde{\mathbf{C}}}_{i}, \quad \widehat{\mathbf{N}}^{B}_i = \begin{bmatrix} \mathbf{0} & \mathbf{0} & \mathbf{D}^B_N \end{bmatrix} \widehat{\tilde{\mathbf{C}}}_{i}$\;
		$\widehat{\mathbf{H}}_i = \mathrm{diag} \left( \left( \mathbf{Y}^B - \widehat{\mathbf{N}}^B_i\right) \widehat{\mathbf{X}}^{A,H}_i \left( \widehat{\mathbf{X}}^{{A}}_i \widehat{\mathbf{X}}^{A,H}_i \right)^{-1} \right)$ (cf. \eqref{eq: problem rtf})
	}
	return $\widehat{\mathbf{X}}^{A}_{I_{\mathrm{se}}}$, $\widehat{\mathbf{H}}_{I_{\mathrm{se}}}$
	\caption{Speech Enhancement Algorithm}
	\label{alg: se}
\end{algorithm}
	
\section{Experiments}
In this section, the performance of the proposed dictionary-based speech enhancement algorithm fusing acoustic and contact microphones for wind noise reduction is compared with several baseline algorithms, which are discussed in Section~\ref{sec: baselines}.
The used dataset and algorithm settings are described in Section~\ref{sec: dataset and settings}.
Finally, the results in terms of objective speech quality and intelligibility metrics are presented in Section~\ref{sec: results}.

\subsection{Baseline Algorithms}
\label{sec: baselines}
In addition to the proposed algorithm, the following baseline algorithms are considered.

First, in order to evaluate the benefit obtained by fusing the acoustic and contact microphone signals, the proposed approach is used \emph{only with the acoustic microphone} spectrogram $\mathbf{Y}^A$ and noise dictionary $\mathbf{D}^A_N$, which is equivalent to setting $\tilde{\mathbf{Y}} = \mathbf{Y}^A$, $\tilde{\mathbf{D}} = \begin{bmatrix} \mathbf{D}^A_X & \mathbf{D}^A_N \end{bmatrix}$, $\tilde{\mathbf{C}} = \begin{bmatrix} \mathbf{C}^{A,T}_X & \mathbf{C}^{A,T}_N \end{bmatrix}{}^T$
%\begin{align}
%	\tilde{\mathbf{Y}} = \mathbf{Y}^A, \quad \tilde{\mathbf{D}} = \begin{bmatrix} \mathbf{D}^A_X & \mathbf{D}^A_N \end{bmatrix}, \quad \tilde{\mathbf{C}} = \begin{bmatrix} \mathbf{C}^{A}_X\\\mathbf{C}^{A}_N \end{bmatrix}
%\end{align}
in \eqref{eq: problem se}, and which can be understood as separating the noisy acoustic microphone signals into speech and noise components, circumventing the need of estimating an \ac{RTF}.
This approach exhibits similarities to previous studies such as in \cite{sigg_speech_2012,ji_speech_2019}, which share the idea of employing both speech and noise dictionaries, but differ mainly by deriving a real-valued gain from the estimated speech and noise components instead of using the estimated speech directly.

Second, the proposed approach is used \emph{only with the contact microphone} spectrogram $\mathbf{Y}^B$ and noise dictionary $\mathbf{D}^B_N$, which is equivalent to setting 
$\tilde{\mathbf{Y}} = \mathbf{Y}^B, \quad \tilde{\mathbf{D}} = \begin{bmatrix} \mathbf{D}^A_X & \mathbf{D}^B_N \end{bmatrix}$, $\tilde{\mathbf{C}} = \begin{bmatrix} \mathbf{C}^{A,T}_X & \mathbf{C}^{B,T}_N \end{bmatrix}{}^T$
%\begin{align}
%	\tilde{\mathbf{Y}} = \mathbf{Y}^B, \quad \tilde{\mathbf{D}} = \begin{bmatrix} \mathbf{D}^A_X & \mathbf{D}^B_N \end{bmatrix}, \quad \tilde{\mathbf{C}} = \begin{bmatrix} \mathbf{C}^{A}_X\\\mathbf{C}^{B}_N \end{bmatrix}
%\end{align}
in \eqref{eq: problem se}, and which can be understood as separating the noisy contact microphone signals into a speech component transformed using the estimated \ac{RTF} and a noise component.

Third, as a traditional algorithm to combine multiple microphone signals by exploiting spatial correlations, a (stationary) \ac{MVDR} beamformer~\cite{gannot_consolidated_2017,van_veen_beamforming:_1988} is considered.
This beamformer aims at minimizing the output noise power while preserving the target speech as described by its \ac{RTF} vector, i.e.,
\begin{align}
	\mathbf{w}_{\mathrm{MVDR}}(f) = \frac{\widehat{\boldsymbol{\Phi}}_n^{-1}(f) \widehat{\mathbf{h}}_0(f)}{\widehat{\mathbf{h}}^H_0(f) \widehat{\boldsymbol{\Phi}}_n^{-1}(f) \widehat{\mathbf{h}}_0(f)},
\end{align}
where $\boldsymbol{\Phi}_n(f)$ denotes the estimated noise spatial covariance matrix $\mathcal{E}\left( \mathbf{n}(f,t) \mathbf{n}^H(f,t) \right)$ with $\mathbf{n}(f,t) = \begin{bmatrix} N^A(f,t) & N^B(f,t) \end{bmatrix}^T$, $\widehat{\mathbf{h}}_0(f) = \begin{bmatrix}	1 & \widehat{H}_0(f) \end{bmatrix}^T$ denotes the \ac{RTF} vector estimate, and $\widehat{X}^A_{\mathrm{MVDR}}(f,t) = \mathbf{w}_\mathrm{MVDR}^H(f) \begin{bmatrix} Y^A(f,t)&Y^B(f,t) \end{bmatrix}^T$.
As such, contrary to the dictionary-based algorithms, the beamformer does not make use of the spectral structure of the speech component.

Fourth, the color correction algorithm uses only the contact microphone signal $Y^B(f,t)$ as an input and aims at equalizing the \ac{RTF} between the speech components at the acoustic and contact microphones, i.e., $\widehat{X}^A(f,t) = \widehat{H}_0^{-1}(f) Y^B(f,t)$, assuming that the \ac{SNR} at the contact microphone is higher.

\subsection{Dataset and Settings}
\label{sec: dataset and settings}
%To train the speech and noise dictionaries needed for the proposed speech enhancement algorithm, speech data as well as noise data characterizing the noise components at the acoustic and contact microphones is required.
To train the speech dictionary $\mathbf{D}^A_X$, we used \SI{10}{\minute} of randomly chosen clean speech utterances from the \texttt{dev-clean} subset of the LibriTTS dataset~\cite{zen_libritts_2019}.
To train the noise dictionaries $\mathbf{D}^A_N$ and $\mathbf{D}^B_N$, we used \SI{2}{\minute} of fan noise recorded by the acoustic and contact microphones at different wind speeds (\SI[per-mode=symbol]{3}{\meter\per\second} and \SI[per-mode=symbol]{5}{\meter\per\second}).
Each of the trained dictionaries consists of $N=1000$ atoms, such that the composite dictionary $\tilde{\mathbf{D}}$ in \eqref{eq: composites definition} consists of $3N=3000$ atoms.
For testing, we considered clean speech from two stationary female and male speakers each as well as fan noise at \SI[per-mode=symbol]{3}{\meter\per\second} and \SI[per-mode=symbol]{5}{\meter\per\second} recorded by the acoustic and contact microphones, with a constant fan speed per utterance.
Different fan noise segments were chosen for the train and test datasets.
Speech and noise components were mixed at broadband \acp{SNR} from \SIrange{-10}{5}{\decibel} (computed at the acoustic microphone), resulting in 10 utterances with a length of \SI{4}{\second} each per \ac{SNR} condition.
All data was sampled at \SI{16}{\kilo\hertz}.
For the \ac{STFT}, we used square root Hann windows for analysis and synthesis with a frame length of \SI{32}{\milli\second} and an overlap of \SI{50}{\percent}.
Based on preliminary experiments, we used the microphone weight $\eta=0.4$ in \eqref{eq: composites definition}, placing more weight on the contact microphone than on the acoustic microphone.
To obtain the \ac{RTF} estimate $\widehat{\mathbf{H}}_0$ required for the initialization of Alg.~\ref{alg: se} and the \ac{MVDR} and color correction algorithms, we used the \ac{CW} method~\cite{markovich-golan_multichannel_2009}, where an energy-based voice activity detector~\cite{shokouhi_idnavidpy_vad_tool_2021} was used to estimate the noise spatial covariance matrix $\widehat{\boldsymbol{\Phi}}_n(f)$.
The maximum numbers of iterations in Algs. \ref{alg: dl}, \ref{alg: sc}, and \ref{alg: se} were set to $I_{\mathrm{dict}}=25$, $I_{\mathrm{sc}}=1000$, and $I_{\mathrm{se}}=5$, respectively.

\subsection{Results}
\label{sec: results}
Speech enhancement performance is measured as the average improvements in terms of \ac{PESQ}~\cite{rix_2001_perceptual} and \ac{STOI}~\cite{taal_short-time_2010} w.r.t. the noisy acoustic microphone signals, using the clean speech at the acoustic microphone as the reference signal.

To investigate the sensitivity of the proposed dictionary-based algorithms w.r.t. the sparsity weight $\lambda$ in \eqref{eq: composite cost}, Fig.~\ref{fig:results: lambdas} shows the average \ac{PESQ} and \ac{STOI} improvements and standard deviations using only the acoustic microphone, only the contact microphone, or using both microphones for three different values of $\lambda$, where the rule-of-thumb value of $\lambda = F^{-0.5}$ originates from~\cite{mairal_online_2010}.
It can be observed that, for larger values of $\lambda$, the performance tends to decrease when only using the acoustic microphone, increase when only using the contact microphone, and stay approximately the same when using both microphones.
Furthermore, it can be clearly observed that the proposed algorithm fusing both microphones significantly outperforms using only one of the microphones.
For the following experiment, the sparsity weight is fixed to $\lambda=F^{-0.5}$.

For different input \acp{SNR}, Fig.~\ref{fig:results: baselines} shows the average \ac{PESQ} and \ac{STOI} improvements and standard deviations of the proposed dictionary-based algorithms and the considered baseline algorithms.
Comparing the performance of the proposed algorithms using only one microphone, as expected, the simulation results show that the contact microphone is advantageous in low-\ac{SNR} conditions, whereas the acoustic microphone is advantageous in high-\ac{SNR} conditions.
Comparing the proposed approach using only the contact microphone signal with the color correction algorithm, the simulation results show that jointly estimating the speech and noise components as well as the \ac{RTF} is beneficial compared with only utilizing the estimated \ac{RTF}.
Only the proposed dictionary-based microphone algorithm fusing both microphones yields consistent \ac{PESQ} and \ac{STOI} improvements, performing similarly or better than all other considered algorithms. 
	
\section{Conclusion}
In this paper we proposed a dictionary-based microphone fusion algorithm to combine the advantages of acoustic and contact microphones for wind noise reduction.
The algorithm extends conventional single-microphone dictionary-based speech enhancement algorithms by simultaneously modeling the acoustic and contact microphone speech and noise components.
While the relationship of the speech components at both microphones is modeled using an \ac{RTF}, the noise components are modeled using separate noise dictionaries.
%The proposed iterative algorithm alternates between estimating the \ac{RTF} and sparse coding the speech and noise components using pre-trained dictionaries.
Simulation results demonstrate the advantage of fusing the microphone signals using the proposed algorithm compared with using either one of the microphones individually.
%In addition, the proposed dictionary-based microphone fusion algorithm outperforms a number of baseline algorithms, including color correction and stationary \ac{MVDR} beamforming.
\printbibliography

\end{document}